\documentclass[iop]{emulateapj}

\usepackage{graphicx}
\usepackage[toc,page]{appendix}
\usepackage{amssymb}
\usepackage{amsmath}
\usepackage{array}
\newcommand{\degree}{\ensuremath{^\circ}}

\newcommand{\eg}{e.g., }

\newcommand{\etal}{et al.}

\mathchardef\mhyphen="2D 
\newcommand\norot{\mathop{no \mhyphen rot}}
\newcommand{\code}[1]{\texttt{#1}} 

\newcommand{\bsym}[1]{\mbox{\boldmath $#1$}} 
\newcommand{\cross}{\bsym{\times}}           
\begin{document}

\title{ROTATION-DEPENDENT CATASTROPHIC DISRUPTION OF GRAVITATIONAL AGGREGATES}
\author{Ronald-Louis Ballouz$^1$, Derek C. Richardson$^1$, Patrick Michel$^2$, and Stephen R. Schwartz$^{2,1}$}
\affil{$^1$Department of Astronomy, University of Maryland, College Park, MD, 20742-2421, USA; rballouz@astro.umd.edu\\ 
$^2$Lagrange Laboratory, University of Nice Sophia Antipolis, CNRS, Observatoire de la C\^{o}te d'Azur, C.S.\ 34229, F-06304 Nice Cedex 4, France}

\submitted{Received 2014 January 30; accepted 2014 May 30}
\begin{abstract}
We carry out a systematic exploration of the effect of pre-impact rotation on the outcomes of low-speed collisions between planetesimals modeled as gravitational aggregates. We use \code{pkdgrav}, a cosmology code adapted to collisional problems and recently enhanced with a new soft-sphere collision algorithm that includes more realistic contact forces. A rotating body has lower effective surface gravity than a non-rotating one and therefore might suffer more mass loss as the result of a collision. What is less well understood, however, is whether rotation systematically increases mass loss on average regardless of the impact trajectory. This has important implications for the efficiency of planet formation via planetesimal growth, and also more generally for the determination of the impact energy threshold for catastrophic disruption (leading to the largest remnant retaining 50\% of the original mass), as this has generally only been evaluated for non-spinning bodies. We find that for most collision scenarios, rotation lowers the threshold energy for catastrophic dispersal. For head-on collisions, we develop a semi-analytic description of the change in the threshold description as a function of the target's pre-impact rotation rate, and find that these results are consistent with the ``universal law'' of catastrophic disruption developed by Leinhardt \& Stewart. Using this approach, we introduce re-scaled catastrophic disruption variables that take into account the interacting mass fraction of the target and the projectile in order to translate oblique impacts into equivalent head-on collisions.  
\keywords{minor planets, asteroids: general - planets and satellites: formation -  planets and satellites: physical evolution }
\end{abstract}

\section{Introduction}

\indent Much of the evolution of small solar system bodies (SSSBs) is dominated by collisions, whether from the initial build-up of planetesimals (Lissauer 1993) or the subsequent impacts between remnant bodies that exist today (e.g., Michel et al.\ 2004). Outcomes of collisions between SSSBs are divided into two regimes: those dominated by material strength and those dominated by self-gravity (Holsapple 1994). The transition from the strength to the gravity regime may occur at body sizes as small as a few kilometers or less for basalt (Benz \& Asphaug 1999; Jutzi et al.\ 2010). After their formation, planetesimals interacted with one another in a dynamically cold disk (Levison et al.\ 2010). This allowed planet-size objects to form through collisonal growth.\\ 
\indent Since the dominant source of confining pressure for planetesimal-size SSSBs is self-gravity, rather than material strength, they can be assumed to be gravitational aggregates (Richardson et al. 2002). Hence, the collisions can often be treated as impacts between rubble piles, the outcomes of which are dictated by collisional dissipation parameters and gravity (Leinhardt et al.\ 2000; Leinhardt \& Richardson 2002). Understanding the effects that contribute to changes in the mass (accretion or erosion) of gravitational aggregates is important for collisional evolution models of the early solar system (e.g., Leinhardt \& Richardson 2005; Weidenschilling 2011). The outcomes of impacts in these models are parameterized through a catastrophic disruption threshold $Q^{\star}_{D}$ (e.g., Benz \& Asphaug 1999), which is the specific impact energy required to gravitationally disperse half the total mass of the system, such that the largest remnant retains the other half of the system mass. However, few studies have accounted for the effect of pre-impact rotation on the size evolution of SSSBs.\\
\indent A rotating body has lower effective surface gravity than a non-rotating one (with the difference being greatest for surface material at the equator and decreasing for material clFig.oser to the rotation axis). Therefore, a rotating body might suffer more mass loss as the result of a collision. A recent laboratory study by Morris et al.\ (2012) suggests this is true for solid bodies. What is less well understood, however, is whether rotation systematically increases mass loss on average regardless of the impact trajectory.\\
\indent In order to explain the collisional evolution of rotation rates of asteroids, Dobrovolskis and Burns (1984) evaluated analytically the sensitivity of mass loss to rotation for cratering impacts on rigid bodies. They found that the angle-averaged mass loss for cases with rotation is enhanced by factors of $\sim$10\%--40\% compared to cases without rotation for rotation speeds $\sim$ 40--80\% of the critical spin rate (see their Figure 2). Analytic and numerical work by Cellino et al.\ (1990) showed that catastrophic disruptions, rather than cratering events, were a bigger contributor to the rotational evolution of asteroids through an angular momentum ``splash'' process; however, they and subsequent authors (e.g., Love \& Ahrens 1996) focused on the effects of spin-state evolution change rather than mass loss.\\
\indent Other authors have included pre-impact rotation in their numerical simulations of planetesimal and protoplanet collisions; however, except for Takeda \& Ohtsuki (2009; see below), none have systematically studied its contribution to mass loss in the dispersive regime. Using a hard-sphere model, Leinhardt et al.\ (2000) performed numerical simulations of collisions of equal-size bodies with pre-impact rotation; however, their work focused on the effect of rotation on the shape of the largest remnant. Canup (2008) studied the effect of pre-impact rotation on lunar formation; however, the work focused on a non-dispersive collision regime. Using a soft-sphere collision code, Takeda \& Ohtsuki (2009) performed simulations of hyper-velocity impacts on rotating $\sim 10$ km size bodies. They found that mass loss is only sensitive to rotation when the target has an in initial spin period close to break-up; otherwise, the collisional energy needed to disrupt a rubble-pile object is not affected by initial rotation. They argued that, upon collision, the ejection speeds of fragments in the hemisphere rotating away from the projectile (prograde direction) are accelerated by the initial rotation, but this is balanced by fragments in the hemisphere rotating toward the projectile (retrograde direction) being decelerated. However, their analysis was restricted to targets with initial rotations of 2.6 and 4.6 revolutions per day (9.23 and 5.58 hr, respectively). Furthermore, their work focused on the efficiency of angular momentum transfer in catastrophic collisions.\\
\indent In this paper, we expand upon the work of previous authors by performing a systematic study of the effect of pre-impact rotation on the energy required to disperse material from km-size gravitational aggregates rotating with spin periods of 3, 4.5, and 6 hr. We solve numerically the outcomes of rubble-pile collisions using a combination of a soft-sphere discrete element method (SSDEM) collisional code and a numerical gravity solver, \code{pkdgrav} (Stadel 2001), which is needed to accurately model the reaccumulation stage. SSDEM has the numerical resolution to determine the mechanics involved in enhancing or diminishing the amount of mass loss associated with collisions onto a rotating target. SSDEM permits realistic modeling of multi-contact and frictional forces between discrete indestructible particles. Thus, it is well suited to study low-speed (a few to tens of m s$^{-1}$) impacts, as it can model robustly collisions that do not produce irreversible shock damage to material (as in hypervelocity, km s$^{-1}$ impacts). In the quasi-steady-state collisional system generally present in a protoplanetary disk, impact speeds are typically of order the escape speed of the largest body in the vicinity. Until the largest body becomes protoplanet sized, impacts will be typically at speeds less than the sound speed of the assumed rocky material. Hence, we limit our study to collisions that occur at subsonic speeds. Most significant collisions today occur at supersonic speeds; however, studies of supersonic collisions require the use of shock physics codes, which include the effects of irreversible shock deformation. \\
\indent Furthermore, we attempt to revise the dependence of catastrophic disruption on the impact parameter $b = \sin\theta$, where $\theta$ is the angle between the projectile's path and the target's center at impact (see Section 2.3). Previous studies (Canup 2008; Leinhardt \& Stewart 2012) have shown that the increase in the threshold for catastrophic disruption for oblique impacts is due to a reduction in interacting projectile mass. These authors provide a formulation parameterized by the fraction of interacting mass, $\alpha$. We show that this does not account adequately for the increase in the catastrophic dispersal threshold for impacts with $b\ne0$ but $\alpha \sim 1$ (impacts where most of the projectile interacts with the target). We discuss a possible revision to the formulation of the catastrophic disruption variables that includes the effective interacting target material. By only taking into account the mass of material that interacts in the collision, oblique impacts are rescaled into equivalent head-on collisions such that they are well described by the so-called ``universal'' law for catastrophic disruption (Leinhardt \& Stewart 2012).\\
\indent Our results have important implications for the efficiency of planet formation via planetesimal growth, and for the determination of the impact energy threshold for catastrophic disruption, as this has generally only been evaluated for non-spinning bodies. In Section 2 we explain the computational methods and outline the parameter space that we explore. In Section 3 we provide our results. In Section 4 we discuss these results in the context of the ``universal'' law for catastrophic disruption and formulate a semi-analytic description of the dependence of catastrophic disruption on pre-impact rotation. We summarize and offer perspectives in Section 5. 

\section{Methodology}

\subsection{Numerical Method}

We use \code{pkdgrav}, a parallel $N$-body gravity tree code (Stadel 2001) adapted for particle collisions (Richardson \etal\ 2000; 2009; 2011). Originally collisions in \code{pkdgrav} were treated as idealized single-point-of-contact impacts between rigid spheres. A soft-sphere option was added recently (Schwartz \etal\ 2012); with this option, particle contacts can last many time steps, with reaction forces dependent on the degree of overlap (a proxy for surface deformation) and contact history.\\
\indent The spring/dash-pot model used in \code{pkdgrav}'s soft-sphere implementation is described fully in Schwartz \etal\ (2012). A spherical particle overlapping with a neighbor feels a reaction force in the normal and tangential directions determined by spring constants ($k_n$, $k_t$), with optional damping and effects that impose static, rolling, and/or twisting friction. The damping parameters ($C_n$, $C_t$) are related to the conventional normal and tangential coefficients of restitution used in hard-sphere implementations, $\varepsilon_n$ and $\varepsilon_t$. The static, rolling, and twisting friction components are parameterized by dimensionless coefficients $\mu_s$, $\mu_r$, and $\mu_t$, respectively. Careful consideration of the soft-sphere parameters is needed to ensure internal consistency, particularly with the choice of $k_n$, $k_t$, and time step. The numerical approach has been validated through comparison with laboratory experiments; \eg Schwartz \etal\ (2012) demonstrated that \code{pkdgrav} correctly reproduces experiments of granular flow through cylindrical hoppers, specifically the flow rate as a function of aperture size, Schwartz \etal\ (2013) demonstrated successful simulation of laboratory impact experiments into sintered glass beads using a cohesion model coupled with the soft-sphere code in \code{pkdgrav}, and Schwartz et al.\ (2014) applied the code to low-speed impacts into regolith in order to test asteroid sampling mechanism design.

\subsection{Rubble-pile Model}
Our simulations consist of two bodies with a mass ratio of $\sim 1:10$: a stationary target with mass $M_{\mathrm{targ}}$ and a projectile with mass $M_{\mathrm{proj}}$ ($M_{\mathrm{proj}}$=0.1$M_{\mathrm{targ}}$ for this work) which impacts the target at a speed of $v_\mathrm{{imp}}$. Both the target and projectile are gravitational aggregates of many particles bound together by self-gravity. The particles themselves are indestructible and have a fixed mass and radius. In the simulations reported here, the only friction that is modeled is static friction, for which we assume $\mu_s = 0.5$, corresponding to an internal angle of friction of $\tan^{-1}(\mu_s) \sim 27\degree$ (we discuss the possible outcome dependence on SSDEM parameters in Section $5$).\\
\indent The rubble piles are created by placing equal-sized particles randomly in a spherical cloud and allowing the cloud to collapse under its own gravity with highly inelastic particle collisions. Randomizing the internal structure of the rubble piles reduces artificial outcomes due to the crystalline structure of hexagonal close packing (Leinhardt et al.\ 2000; Leinhardt \& Richardson 2002). Due to symmetry lines and planes in crystalline packing, there is a dependency of the collision outcome on the initial orientation of the target's principal axes. To test the dependence of the collision outcome on initial orientation for a spherically collapsed rubble pile, a series of simulations was performed where the simulation parameters were kept constant except for the initial orientation of the target's equatorial principal axes, which were varied by increments of $45\degree$ about its polar axis. The results of these simulations show that, for a spherically collapsed rubble pile, the dependence of collision outcome on initial orientation is small (mass loss deviations of less than $1\%$ from the mean).\\
\indent For the simulations presented here, the target had an average radius of $R_\mathrm{{targ}} \sim 1.0$ km and bulk density of $\rho_\mathrm{{targ}} \sim  2$ g cm$^{-3}$. The projectile had an average radius of $R_\mathrm{{proj}} \sim 0.5$ km and bulk density of $\rho_\mathrm{{proj}}\sim 2$ g cm$^{-3}$. In order to determine accurately the physical properties (size, shape, mass, angular momentum) of the target after the collision, the rubble piles were constructed with a relatively high number of particles ($N_{\mathrm{targ}} = 10^4, N_{\mathrm{proj}} = 10^3$).\\
\indent The collisional properties of the constituent particles are specified prior to each simulation. These values were fixed at $\epsilon_n=0.8 $ (mostly elastic collisions with some dissipation) and $\epsilon_t=1.0$ (no sliding friction). Furthermore, since SSDEM models treat particle collisions as reactions of springs due to particle overlaps, the magnitude of the normal and tangential restoring forces are determined by the spring constants $k_n$ and $k_t \sim \frac{2}{7}k_n$. We choose $k_n$ by requiring the maximum fractional particle overlap, $x_{\mathrm{max}}$, to be $\sim 1\%$. For rubble-pile collisions, the value of $k_{n}$ can be estimated by:
\begin{equation} k_n \sim m \left(\frac{v_{\mathrm{max}}}{x_{\mathrm{max}}}\right)^{2},\end{equation}
\noindent where $m$ corresponds to the typical mass of the most energetic particles, and $v_{\mathrm{max}}$ is the maximum expected speed in the simulation (Schwartz et al.\ 2012). Thus, for our rubble-pile collisions with speeds $\le 10$ m s$^{-1}$, $k_n \sim 4 \times 10^{11}$ kg s$^{-2}$. The initial separation of the projectile and target, $d$, for all cases was $\sim 4R_{\mathrm{targ}}$, far enough apart that initial tidal effects were negligible. In order for the post-collision system to reach a steady state, the total run-time was set to $\sim 3\times$ the dynamical time for the system, $1/\sqrt{G\rho_{\mathrm{targ}}} \sim 2$ hr. Furthermore, a time step $\Delta t \sim 3$ ms was chosen on the basis of the time required to sample particle overlaps adequately, for the choice of $k_n$ and $x_{\mathrm{max}}$ given above.

\begin{figure}[h!]
	\centering
  \includegraphics[width=0.25\textwidth]{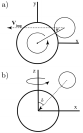}
\caption{\footnotesize{Schematic of two collision scenarios. Panel (a) shows a target impacted in its equatorial plane by a projectile moving right to left with speed $v_{\mathrm{imp}}$. The impact angle $\theta$ is the angle between the line connecting the centers of two bodies and the projectile's velocity vector, at the time of contact. Panel (b) shows a projectile that impacts a target at an angle $\delta$ between the rotation axis ($z$) and the projectile's velocity vector.} }\label{fig:diagrams}
\end{figure}

\subsection{Simulation Parameters and Collision Geometries}
In order to probe the effect of rotation on collision outcome, simulations with the target rubble pile having an initial spin period $P_{\mathrm{spin}}$ of 3, 4.5, and 6 hr (values well above the spin break-up limit for a rubble pile of bulk density $\sim 2$ g cm$^{-3}$, $\sim 2.3$ hr), were compared against runs with the target having no initial spin. For the material parameters assumed here, a rotating spherical rubble-pile would likely find a new spin-shape equilibrium. However, the impacts of our simulations occur quickly enough that the target does not deviate from its spherical shape before disrupting. Determining the effect of pre-impact shape on collision outcome is outside the scope of this study and is left for future work. For every spin period, simulations were done with a range of impact speeds such that there was adequate coverage of the gravitational dispersal regime (collisions that result in a system losing $0.1$--$0.9$ times its total mass).\\
\indent Furthermore, three different collision geometries were explored in this work, each of which depended on two different collision parameters. The first was the impact parameter $b = \sin \theta$, where $\theta$ is the angle between the line connecting the centers of two bodies and the projectile's velocity vector (see Figure 1(a)). The second parameter was the angle $\delta$, which is the angle between the target's rotation axis and the projectile's velocity vector (see Figure 1(b)). In this study, the effect of each parameter on the collision outcome was studied separately and compared against the standard case of a head-on collision. In a head-on collision, the projectile's velocity vector is normal to the target's rotation axis and is directed towards its center ($b = 0$ and $\delta = 90\degree$).\\
\indent For oblique impacts, the impact parameter $b \neq 0$. The impact parameter has a significant effect on the collision outcome because the total mass of the projectile may not completely intersect the target when the impact is oblique (e.g., Yanagisawa \& Hasegawa 2000; Canup 2008; Leinhardt \& Stewart 2012). Thus, when the projectile is large enough compared to the target, a portion of the projectile may shear off and only the kinetic energy of the interacting fraction of the projectile will be involved in disrupting the target. For any given collision speed, an oblique impact will erode less mass than a head-on collision. In this study, four values of $b$ were used, $\pm 0.5$ and $\pm 0.7$. For $b<0$, the projectile impacts the target on the hemisphere that rotates towards the projectile, which we define as the retrograde hemisphere. For $b>0$, the projectile impacts the target on the hemisphere that is rotating away from the projectile, the prograde hemisphere (see Figure 1(a)). Hence, if the collision outcome is sensitive to initial rotation, then it is expected that the sign of $b$ will also affect the amount of mass that is dispersed. \\
\indent For non-equatorial impacts, the polar angle $\delta < 90\degree$ (see Figure 1(b)). If the collision outcome is sensitive to the target's pre-impact rotation, then material from the target's equator may preferentially be dispersed due to its lower specific binding energy. However, it is uncertain whether the projectile more efficiently transfers its energy to the target's equator or to its poles upon impact. Hence, this study tests the effect of three different polar impact angles: $\delta = 90\degree$ (collisions directed at the target's equator), $\delta = 45\degree$, and $\delta=0\degree$ (collisions directed at the target's pole). In reality, most collisions will have a combination of non-zero values for both $b$ and $\delta$. 

\section{Results}
The collision of two rubble-pile objects typically results in either net accretion, where the largest remnant has a net gain in mass compared to the mass of the target, or net erosion, where the target has lost mass. Alternatively, a collision could result in no appreciable net accretion or erosion (Leinhardt \& Stewart 2012). These latter types of collisions, called hit-and-run events, typically occur for grazing impacts that have an impact parameter, $|b|$, that is greater than a critical impact parameter $b_{\mathrm{crit}}$ (Asphaug 2010), where $b_{\mathrm{crit}} = R_\mathrm{{targ}}/(R_{\mathrm{proj}}+R_\mathrm{{targ}})$. In this paper, we focus on the dispersive regime, where impact velocities, $v_{\mathrm{imp}}$, are greater than the escape speed from the surface of the target, $v_{\mathrm{esc}}$ (assuming no rotation). The impact speeds in our simulation range from $4$--$30$ $v_{\mathrm{esc}}$, where $v_{\mathrm{esc}} \sim 1$ m s$^{-1}$ is the escape speed from a spherical object with mass $M_{\mathrm{tot}}=M_{\mathrm{proj}}+M_\mathrm{{targ}}$ and density $\rho_{1} = 1 $ g cm$^{-3}$. At impact speeds of $4$--$30$ $v_{\mathrm{esc}}$, the mass of the largest remnant in each simulation, $M_\mathrm{{LR}}$, ranges between $0.2$--$0.8$ $M_{\mathrm{tot}}$. The amount of mass loss at the end of a simulation is found by measuring the final mass of the largest remnant and all material gravitationally bound to it (material with instantaneous orbital energy $\leq 0$). Furthermore, we analyze the mechanics behind rotation-dependent mass loss by comparing the number of escaping particles that originate from different regions of the target. The result of each simulation is summarized in Table 1.\\
{\renewcommand{\arraystretch}{1.2}
\begin{table*}[h]
\caption{\footnotesize{Summary of Collision Geomteries and Mass Loss Outcomes.}}
\label{t:summary_results}
\parbox{.45\linewidth}{\centering
\begin{tabular}{cccccc}
\hline
\hline
$b$ & $\delta$ ($\degree$) & $P_\mathrm{{spin}}$ (h) & $v_{\mathrm{imp}}$ (m s$^{-1}$) & $M_{\mathrm{LR}}/M_\mathrm{{tot}}$ \\
\hline
0 & 90 & $\infty$ & 6.0 & 0.831089\\
0 & 90 & $\infty$ & 7.0 & 0.764666\\
0 & 90 & $\infty$ & 8.0 & 0.683734\\
0 & 90 & $\infty$ & 9.0 & 0.607632\\
0 & 90 & $\infty$ & 10.0 & 0.517345\\
0 & 90 & $\infty$ & 11.0 & 0.428664\\
\hline
0 & 90 & 6 & 6.0 & 0.811708\\
0 & 90 & 6 & 7.0 & 0.742641\\
0 & 90 & 6 & 8.0 & 0.661843\\
0 & 90 & 6 & 9.0 & 0.579304\\
0 & 90 & 6 & 10.0 & 0.494084\\
0 & 90 & 6 & 11.0 & 0.389184\\
\hline
0 & 90 & 4.5 & 6.0 & 0.800477\\
0 & 90 & 4.5 & 7.0 & 0.728418\\
0 & 90 & 4.5 & 8.0 & 0.649403\\
0 & 90 & 4.5 & 9.0 & 0.567309\\
0 & 90 & 4.5 & 10.0 & 0.464814\\
0 & 90 & 4.5 & 11.0 & 0.351837\\
\hline
0 & 90 & 3 & 6.0 & 0.760627\\
0 & 90 & 3 & 7.0 & 0.696922\\
0 & 90 & 3 & 8.0 & 0.599811\\
0 & 90 & 3 & 9.0 & 0.514277\\
0 & 90 & 3 & 10.0 & 0.421599\\
0 & 90 & 3 & 11.0 & 0.305056\\
\hline
0 & 45 & $\infty$ & 5.0 & 0.876133\\
0 & 45 & $\infty$ & 6.0 & 0.807526\\
0 & 45 & $\infty$ & 7.0 & 0.739923\\
0 & 45 & $\infty$ & 8.0 & 0.652526\\
0 & 45 & $\infty$ & 9.0 & 0.575735\\
0 & 45 & $\infty$ & 10.0 & 0.496087\\
\hline
0 & 45 & 6 & 5.0 & 0.877857\\
0 & 45 & 6 & 6.0 & 0.805796\\
0 & 45 & 6 & 7.0 & 0.725621\\
0 & 45 & 6 & 8.0 & 0.634082\\
0 & 45 & 6 & 9.0 & 0.554188\\
0 & 45 & 6 & 10.0 & 0.485355\\
\hline
0 & 45 & 3 & 5.0 & 0.846789\\
0 & 45 & 3 & 6.0 & 0.767449\\
0 & 45 & 3 & 7.0 & 0.684930\\
0 & 45 & 3 & 8.0 & 0.565474\\
0 & 45 & 3 & 9.0 & 0.441618\\
0 & 45 & 3 & 10.0 & 0.378795\\
\hline
0 & 0 & $\infty$ & 5.0 & 0.881670\\
0 & 0 & $\infty$ & 6.0 & 0.817889\\
0 & 0 & $\infty$ & 7.0 & 0.734517\\
0 & 0 & $\infty$ & 8.0 & 0.649638\\
0 & 0 & $\infty$ & 9.0 & 0.562160\\
0 & 0 & $\infty$ & 10.0 & 0.472488\\
\hline
0 & 0 & 6 & 5.0 & 0.874406\\
0 & 0 & 6 & 6.0 & 0.812157\\
0 & 0 & 6 & 7.0 & 0.724058\\
0 & 0 & 6 & 8.0 & 0.627248\\
0 & 0 & 6 & 9.0 & 0.531327\\
0 & 0 & 6 & 10.0 & 0.433020\\
\hline
0 & 0 & 3 & 5.0 & 0.865095\\
0 & 0 & 3 & 6.0 & 0.777822\\
0 & 0 & 3 & 7.0 & 0.684112\\
0 & 0 & 3 & 8.0 & 0.481345\\
0 & 0 & 3 & 9.0 & 0.347406\\
0 & 0 & 3 & 10.0 & 0.206721\\
\hline
\end{tabular}}
\hfill
\parbox{.45\linewidth}{\centering
\begin{tabular}{cccccc}
\hline
\hline
$b$ & $\delta$ ($\degree$) & $P_\mathrm{{spin}}$ (h) & $v_{\mathrm{imp}}$ (m s$^{-1}$) & $M_{\mathrm{LR}}/M_\mathrm{{tot}}$ \\
\hline
+0.5 & 90 & $\infty$ & 5.0 & 0.855201\\
+0.5 & 90 & $\infty$ & 10.0 & 0.694707\\
+0.5 & 90 & $\infty$ & 15.0 & 0.513617\\
+0.5 & 90 & $\infty$ & 20.0 & 0.299089\\
\hline	
+0.5 & 90 & 6 & 5.0 & 0.834180\\
+0.5 & 90 & 6 & 10.0 & 0.671891\\
+0.5 & 90 & 6 & 15.0 & 0.490690\\
+0.5 & 90 & 6 & 20.0 & 0.312836\\
\hline
+0.5 & 90 & 3 & 5.0 & 0.797677\\
+0.5 & 90 & 3 & 10.0 & 0.638241\\
+0.5 & 90 & 3 & 15.0 & 0.445009\\
+0.5 & 90 & 3 & 20.0 & 0.289541\\
\hline
-0.5 & 90 & $\infty$ & 5.0 & 0.852571\\
-0.5 & 90 & $\infty$ & 10.0 & 0.699535\\
-0.5 & 90 & $\infty$ & 15.0 & 0.540372\\
-0.5 & 90 & $\infty$ & 20.0 & 0.327492\\
\hline
-0.5 & 90 & 6 & 5.0 & 0.865060\\
-0.5 & 90 & 6 & 10.0 & 0.702961\\
-0.5 & 90 & 6 & 15.0 & 0.511956\\
-0.5 & 90 & 6 & 20.0 & 0.326386\\
\hline
-0.5 & 90 & 3 & 5.0 & 0.860442\\
-0.5 & 90 & 3 & 10.0 & 0.690634\\
-0.5 & 90 & 3 & 15.0 & 0.481739\\
-0.5 & 90 & 3 & 20.0 & 0.315096\\
\hline
+0.7 & 90 & $\infty$ & 5.0 & 0.876840\\
+0.7 & 90 & $\infty$ & 10.0 & 0.796897\\
+0.7 & 90 & $\infty$ & 15.0 & 0.715336\\
+0.7 & 90 & $\infty$ & 20.0 & 0.619042\\
+0.7 & 90 & $\infty$ & 25.0 & 0.532207\\
+0.7 & 90 & $\infty$ & 30.0 & 0.424618\\
\hline
+0.7 & 90 & 6 & 5.0 & 0.863776\\
+0.7 & 90 & 6 & 10.0 & 0.786793\\
+0.7 & 90 & 6 & 15.0 & 0.695417\\
+0.7 & 90 & 6 & 20.0 & 0.601023\\
+0.7 & 90 & 6 & 25.0 & 0.504266\\
+0.7 & 90 & 6 & 30.0 & 0.414065\\
\hline
+0.7 & 90 & 3 & 5.0 & 0.840861\\
+0.7 & 90 & 3 & 10.0 & 0.750761\\
+0.7 & 90 & 3 & 15.0 & 0.656733\\
+0.7 & 90 & 3 & 20.0 & 0.551418\\
+0.7 & 90 & 3 & 25.0 & 0.462672\\
+0.7 & 90 & 3 & 30.0 & 0.386483\\
\hline
-0.7 & 90 & $\infty$ & 5.0 & 0.874388\\
-0.7 & 90 & $\infty$ & 10.0 & 0.795250\\
-0.7 & 90 & $\infty$ & 15.0 & 0.712159\\
-0.7 & 90 & $\infty$ & 20.0 & 0.628877\\
-0.7 & 90 & $\infty$ & 25.0 & 0.534306\\
-0.7 & 90 & $\infty$ & 30.0 & 0.432360\\
\hline
-0.7 & 90 & 6 & 5.0 & 0.880263\\
-0.7 & 90 & 6 & 10.0 & 0.804069\\
-0.7 & 90 & 6 & 15.0 & 0.720063\\
-0.7 & 90 & 6 & 20.0 & 0.628323\\
-0.7 & 90 & 6 & 25.0 & 0.525654\\
-0.7 & 90 & 6 & 30.0 & 0.390580\\
\hline
-0.7 & 90 & 3 & 5.0 & 0.884772\\
-0.7 & 90 & 3 & 10.0 & 0.787585\\
-0.7 & 90 & 3 & 15.0 & 0.702943\\
-0.7 & 90 & 3 & 20.0 & 0.599283\\
-0.7 & 90 & 3 & 25.0 & 0.477597\\
-0.7 & 90 & 3 & 30.0 & 0.359355\\
\hline
\end{tabular}}
\end{table*}}

\begin{figure}[h!]
\centering
\includegraphics[width=0.5\textwidth]{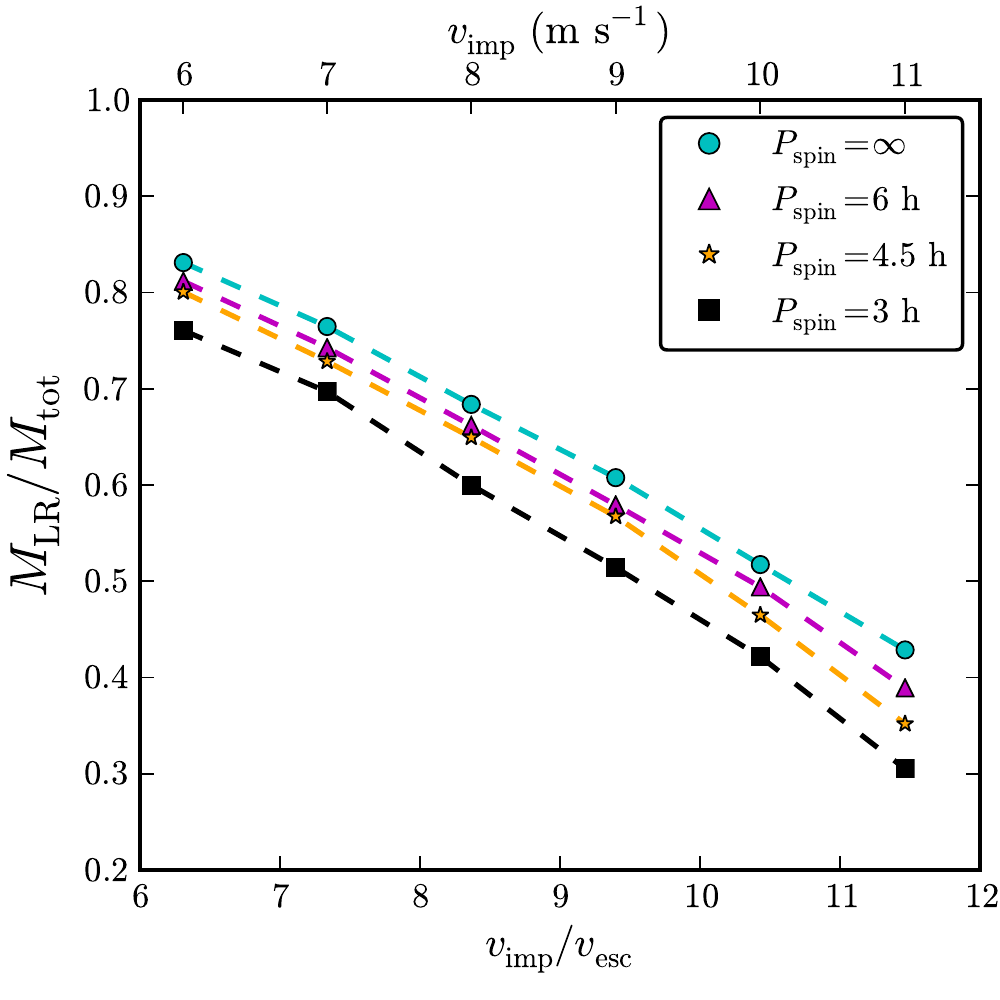} 
\caption{\footnotesize{Mass loss for head-on equatorial impacts ($b=0$, $\delta=90\degree$) is sensitive to pre-impact rotation. The amount of mass that is gravitationally dispersed is proportional to the impact speed. Green triangles, blue stars, and purple squares represent impacts where the target has a pre-impact spin period of 6 hr, 4.5 hr, and 3 hr, respectively. Red-filled circles represent impacts where the target has no pre-impact spin. For head-on equatorial collisions, we derive the dependence of the reduced mass catastrophic disruption threshold, $Q^{\star}_{\mathrm{RD}}$, on the target's pre-impact rotation rate in Section 4.2.}}
\label{f:mesc_headon}
\end{figure}
\subsection{Head-on Equatorial Collisions}
For the nominal case of an equatorial-plane head-on collision, $b=0$ and $\delta=90\degree$. Figure 2 shows that the amount of mass dispersal increases monotonically with collision speed, and that, for head-on equatorial collisions, the amount of mass dispersal is sensitive to initial rotation, as cases with shorter spin periods systematically result in more mass loss. Furthermore, the collision outcomes for the target with an initial period of 6 h approach the outcomes where the rubble pile initially has no spin. Since the spin limit for cohesionless rubble piles of this size and density is $\sim 2.3$ hours, these results are fairly representative of all possible head-on collisions with initial spin (for $\mu_s = 0.5$). Through a simple linear regression of the mass loss as a function of impact speed, we find that the catastrophic disruption threshold $Q^{\star}_{D}$ decreases by a range of $\sim10\%$--$30\%$ for the cases with pre-impact spin studied here. Since the transition from merging to catastrophic disruption may occur over a difference in energy of $\sim 30 \%$ (Leinhardt \& Stewart 2012), our results show that pre-impact spin can play a crucial role in the formation of planetesimals and protoplanets. \\
\begin{figure*}[t!]
\centering
\includegraphics[width=1\textwidth]{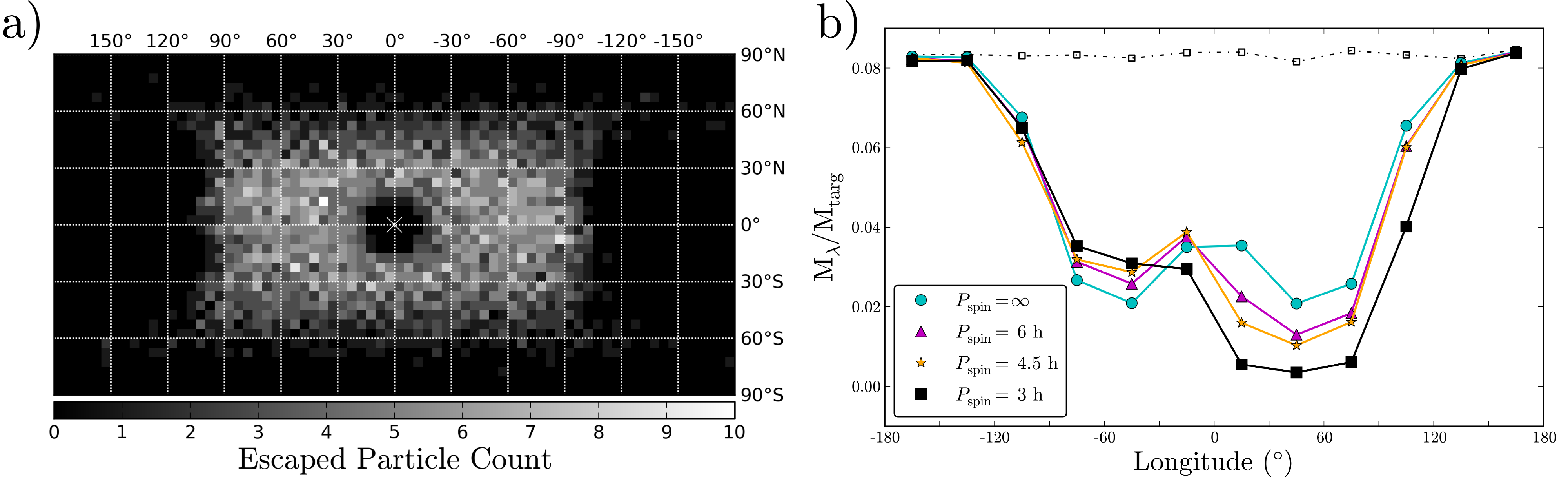}
\caption{\footnotesize{Panel (a) shows a cylindrical equi-distant map projection of the areas of mass loss from the head-on equatorial impact with $v_{\mathrm{imp}}=9$ m s$^{-1}$, and $P_{\mathrm{spin}}=\infty$ (no rotation). For an impactor of finite size, the regions near the impact point [coordinates $(0\degree, 0\degree)$] do not experience mass loss as they are confined between the antipodal material and the incoming projectile during the collision. Rather, a ring of material surrounding the extended impact region is ejected (here shown as a rectangular region due to the distortion of the map projection). Panel (b) shows the provenance of material that make up the largest remnant for simulations with a near-catastrophic impact speed, $v_{\mathrm{imp}}=9$ m s$^{-1}$, and for $P_{\mathrm{spin}}= \infty, 6, 4.5,$ and $3$ hr (cyan circles, magenta triangles, orange stars, and filled black squares, respectively). $\mathrm{M_{\lambda}}$ is the mass in the largest remnant originating from a given longitude range ($30\degree$ bin size) of the target. The open black squares connected by the dash-dotted line represent the initial longitudinal mass distribution of the target. Deviations from a constant value ($\sim 1/12$) are due to the slight asphericity of the target. The sensitivity of mass loss to pre-impact rotation exists due to an enhancement in the amount of escaping material from the prograde hemisphere  ($\mathrm{Longitude} > 0\degree$), which more than compensates for a corresponding greater retention of material from the retrograde hemisphere ($\mathrm{Longitude} < 0\degree$).}}\label{f:escape_map_loss}
\end{figure*}

\indent In order to obtain a better understanding of the underlying mechanics of rotationally enhanced mass loss, we considered the geometrical effects associated with a collision. By tracking the provenance of escaping particles originating from the target, we studied the likelihood of a particle's escape as a function of its initial longitudinal and latitudinal point of origin on the target. Figure 3(a) shows a mass-loss map for the case of a head-on collision with specific impact energy close to catastrophic disruption ($v_{\mathrm{imp}} = 9$ m s$^{-1}$). The collision creates an extended impact region proportional to the projectile's size ($R_{\mathrm{proj}} \sim 0.5$  km), shown in Figure 3(a) as the black region in the middle of the map. Material within this region is retained by the largest remnant as it is enclosed between the incoming projectile and the target's antipodal region (material $180\degree$ from the impact point). The escaping material originates from a nearly symmetrical ring about the impact region. Closer inspection of this escaping material reveals that the enhancement in mass loss is due to a preferential escape of material from prograde (positive) longitudes (Figure 3(b)) (a similar result was found by Takeda \& Ohtsuki (2009)). For this analysis, particles in the initial rubble-pile target were binned into $30 \degree$ longitudinal spherical wedges. Since the target is nearly spherically symmetric, we consider longitudinal bins of equal size. Figure 4 shows the number and sign convention that is used. The $0\degree$ longitude point is defined as the meridian of the target aligned with the impactor's velocity vector at the beginning of the simulation. Figure 3(b) shows the mass distribution of the largest post-impact remnant as a function of the particle origin. For no pre-impact spin, the largest post-collision remnant is composed of a near-equal amount of material from prograde and retrograde hemispheres. This is reflected in the symmetry of the longitudinal mass-distribution profile shown in Figure 3(b) (cyan circles). When a target has pre-impact spin, more material in the retrograde longitudes (negative values) is retained by the largest remnant, while more material located in the prograde longitudes escape. This is due to retrograde material having tangential velocities that are anti-aligned with the impact velocity vector, and prograde material having tangential velocities that are aligned. This can be seen in the increase in the asymmetry of the mass-distribution profiles for increasing spin rate (magenta triangles, orange stars, and black squares in Figure 3(b)). However, there is an imbalance between mass retention and mass escape in opposite hemispheres for cases with pre-impact spin. Relative to the case with no rotation, the troughs of the mass distributions ($30\degree<|\mathrm{Longitude}|<60\degree$) exhibit an extra $\sim20\%--50\%$ of mass retention on the retrograde hemisphere; but, an extra $\sim35\%--85\%$ of mass escapes on the prograde hemisphere. This phenomenon is observed in all cases of pre-impact spin. It is this process that eventually determines the total net enhancement in mass loss as a function of increasing pre-impact spin seen in Figure 2.\\
\indent However, head-on collisions are not the most likely collision geometry; rather, a collsion with $b \sim 0.7$ ($\theta\sim45\degree$) is the most common on average (Love \& Ahrens 1996). Furthermore, the mechanics of mass loss enhancement due to pre-impact spin appears to be linked directly with the manner in which loading on the target occurs. Therefore, we extend our analysis to study the dependence of mass loss with the combined effects of non-zero $b$ and $\delta$ by studying each in isolation.

\begin{figure}[h!]
\centering
\includegraphics[width=0.3\textwidth]{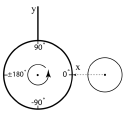} 
\caption{\footnotesize{In order to determine the longitudinal origin of escaping particles, the target is divided into equal-size bins of longitudes. Following the right-hand rule, negative values of the longitude correspond to negative values along the $y$-axis when spin-angular momentum vector is aligned with the positive $z$-axis.} }
\label{f:longitude_diagram}
\end{figure}

\begin{figure*}[t!]
\centering
\includegraphics[width=0.6\textwidth]{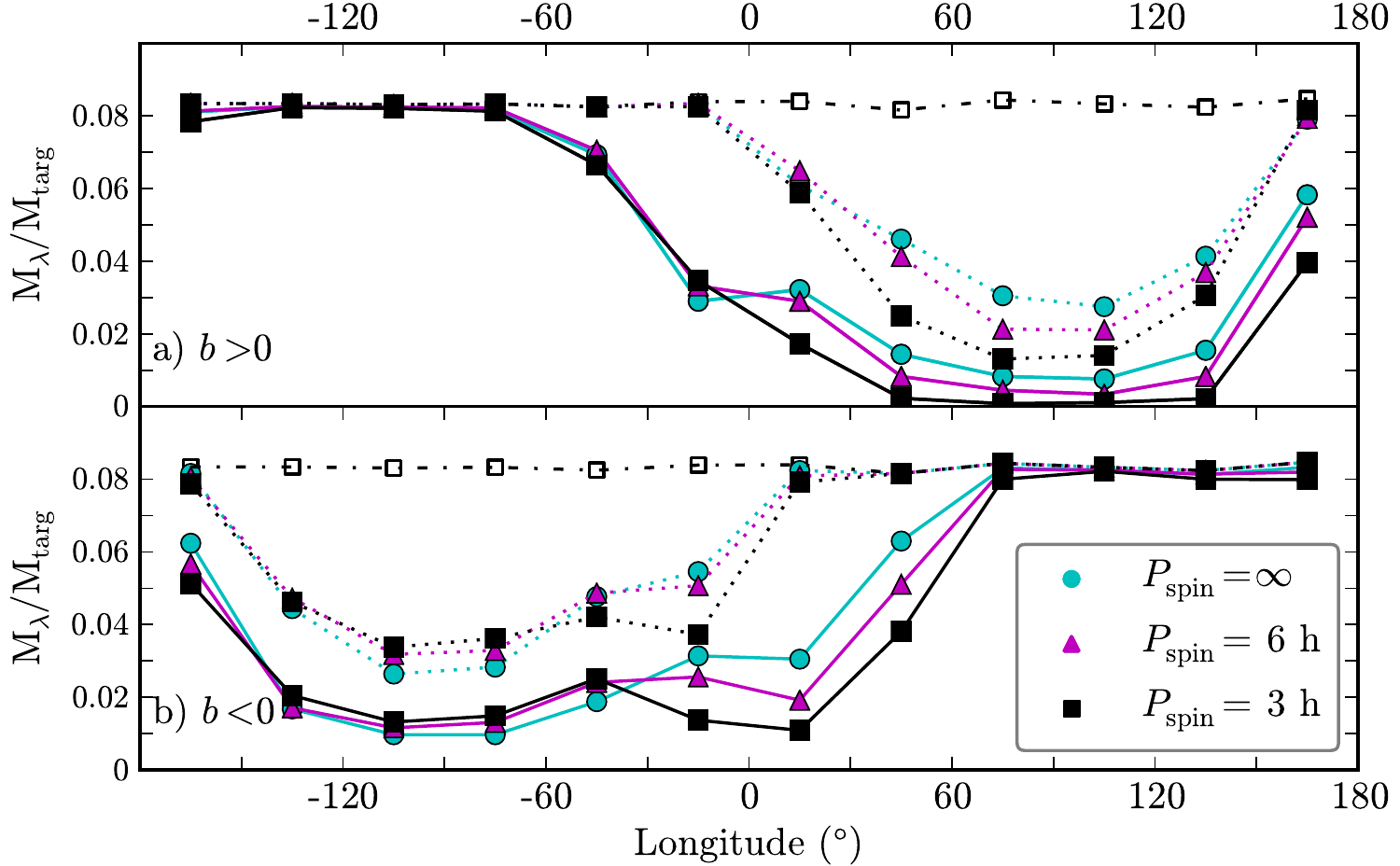} 
\caption{\footnotesize{Mass-loss distribution for oblique equatorial impacts with $v_{\mathrm{imp}} = 15$ m s$^{-1}$. Panel (a) shows the distribution for impacts onto the prograde hemisphere (positive b), where the solid lines are for collisions with $b = + 0.5$, and dotted lines are for $b = + 0.7$. Panel (b) shows the distribution for impacts onto the retrograde hemisphere (negative b), where the solid lines are for collisions with $b = - 0.5$, and dotted lines are for $b = - 0.7$. See text for discussion.}}
\label{f:longitude_b}
\end{figure*}

\subsection{Oblique Equatorial Collisions}

Previous studies have shown that the catastrophic disruption criterion is highly sensitive to the impact parameter (Leinhardt et al.\ 2000; Leinhardt \& Stewart 2012). For oblique impacts, the energy of the projectile may not completely intersect the target. For certain values of $b$, a segment of the projectile may be able to shear off, and, consequently, this material does not interact with the target, effectively lowering the specific impact energy. Hence, a greater impact speed is required to reach catastrophic disruption compared to a head-on collision. Previous studies used a simple geometric model to determine the fraction of interacting projectile mass (Canup 2008; Leinhardt \& Stewart 2012). The revised mass is then used in scaling the catastrophic disruption criteria (Leinhardt \& Stewart 2012).\\ 
\indent We find that mass dispersal is only sensitive to pre-impact rotation when $b > 0$, for the range in $b$ values considered (see Table 1). For $b < 0$, the projectile impacts the target on its retrograde hemisphere; hence, the tangential velocities of rotating particles are either anti-aligned or perpendicular to the impact velocity vector. Figure 5 shows the mass distribution as a function of longitudinal origin for $v_{\mathrm{imp}} = 15$ m s$^{-1}$. The solid and dotted curves are for $b \pm 0.5$ and $b \pm 0.7$, respectively. Figure 5(a) shows that, for collisions onto the prograde hemisphere, pre-impact rotation systematically increases mass loss (less material is retained at each longitude bin). This is similar to the results found for head-on collisions in Section 3.1. However, for the cases with $b=+0.5$ (solid lines), there is a slight inversion in mass retention between spin and no-spin cases retrograde at longitudes (as expected, based on the results from the previous section). This inversion is more apparent in Figure 5(b), which shows that collisions onto the retrograde hemisphere result in a rough balance between mass retention and mass-loss enhancement when pre-impact spin is introduced. Rotation causes increases in mass retention at longitudes westward of the impact point, and an equal decrease at longitudes eastward of the impact point.  Hence, the net effect is that, for retrograde impacts, pre-impact spin does not enhance mass loss very much. Furthermore, for the oblique impacts considered here, the rubble pile is efficient at dissipating the impact energy such that gravitational dispersal is localized to the hemisphere of the target that was impacted, rather than being a global effect.\\

\subsection{Head-on Collisions with $\delta<90\degree$}

\indent Since the effective acceleration of a particle on a rotating body is a function of its colatitude, we would expect that the amount of mass loss a rubble pile experiences is a strong function of the latitude of impact. The effect of varying the latitude of impact was also discussed in Takeda \& Ohtsuki (2009), who studied the effects on the post-impact spin rate. Figure 6 shows the mass dispersal as a function of impact speed for two different values of the polar impact angle, $\delta$. Contrasting with the head-on equatorial case, for low collision speeds, pre-impact rotation does not systematically increase the amount of mass loss. However, for near- and super-catastrophic speeds, the amount of mass loss is greatly enhanced. For such energetic collisions, a target with a $3$ h pre-impact spin period experiences $\sim 25 \%$ (for $\delta=45\degree$) and $\sim 50\%$ ($\delta=0\degree$) more mass loss than a non-spinning target. On average, a pole-on impact onto a spinning target requires $\sim 15\%$ less specific impact energy to reach catastrophic disruption compared to an equatorial head-on impact ($\delta=90\degree$, $b = 0$).  \\
\begin{figure}[h!]
\centering
\includegraphics[width=0.5\textwidth]{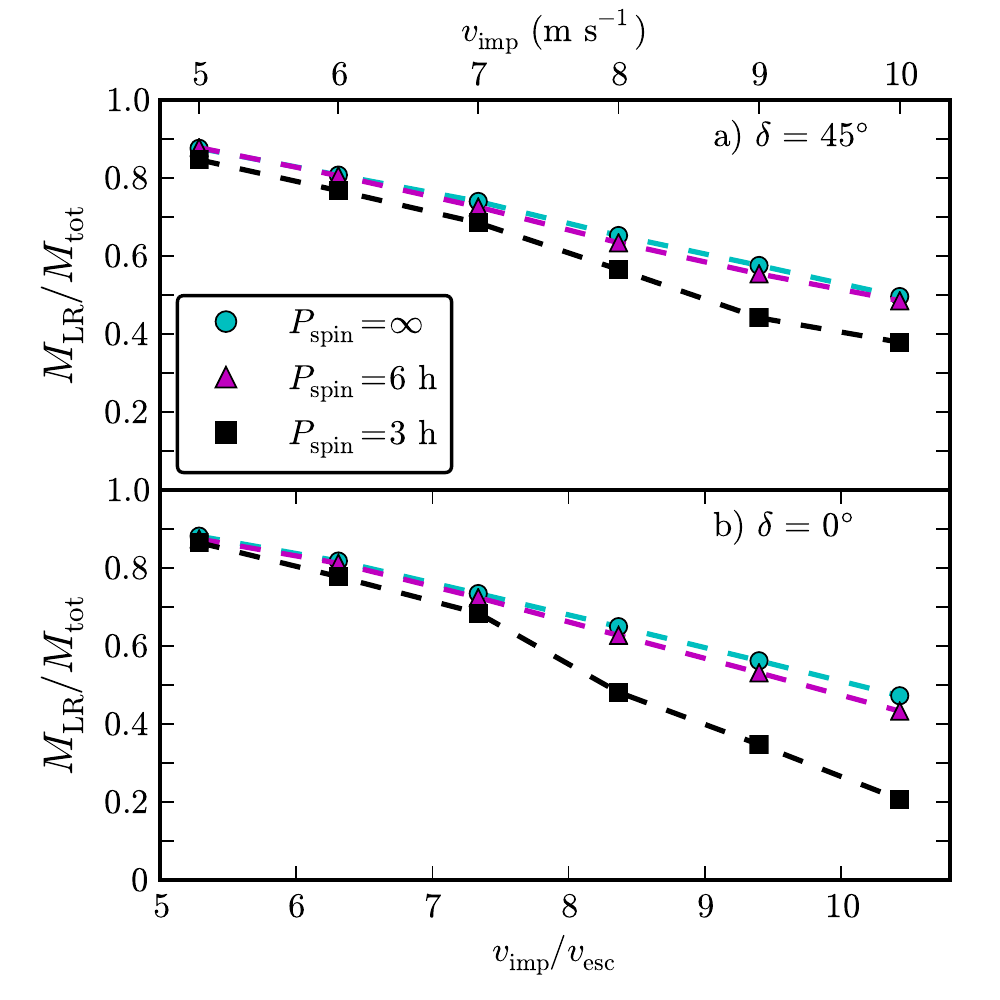} 
\caption{\footnotesize{Mass of the largest remnant as a function of impact speed for cases with $\delta < 90\degree$. Significant enhancements in mass loss only occur for near-catastrophic and super-catastrophic collisions. See text for discussion.}}
\label{f:mesc_delta}
\end{figure}
\indent The enhancement in mass loss for near-polar impacts is due to the escape of equatorial surface particles. As discussed in Section 3.1 and shown in Figure 4(a), the material that escapes originates from a ring about the extended impact region. For the case of pole-on and near-pole-on impacts, this dispersal region extends to the equator, where the effective pre-impact acceleration of particles is greatest. Unlike equatorial impacts ($\delta=90\degree$), material at the equator is not confined by impacting projectile material. Instead, polar material is trapped by the merging projectile material, and the collisional wave disperses material outside the immediate polar region. For highly energetic collisions, this mechanical wave extends to the equator, where particles are unhindered and more readily escape. Hence, for the cases of near- and super-catastrophic collisions, the vertical transfer of impact energy (from the pole to the equator) leads to the ejection of low-latitude particles. Experiencing a lower effective gravitational potential at higher rotation rates, these particles escape more easily.\\
This is demonstrated in Figure 7, which shows the mass distribution of the largest remnant as a function of the absolute latitudinal origin of particles from the target for a pole-on impact ($\delta = 0\degree$). The absolute latitude is used since the target is near-symmetric about the equator. In order to make insightful comparisons, the target's northern and southern hemispheres are sub-divided into six equal-mass regions, which correspond to the following latitude boundaries: $0\degree$, $6\degree$, $13\degree$, $20\degree$, $28\degree$, $39\degree$, and $90\degree$. At low impact speeds, most of the mass loss originates from high latitudes. At sufficiently high impact speeds, more mass loss originates from low latitudes (compare solid cyan curve and dotted black curve in Figure 7(a)). For faster spin rates, mass loss from lower latitudes is even more enhanced (contrast dotted black curves in Figures 7(a) and Figure 7(c)), and the impact speed threshold to move from high-latitude mass-loss to low-latitude mass loss decreases (contrast dot-dashed orange curves in Figure 7(a) and Figure 7(c)). The overall effect is the enhancement in global mass loss at high impact speeds seen in Figure 6. \\

\begin{figure}[h!]
\centering
\includegraphics[width=0.5\textwidth]{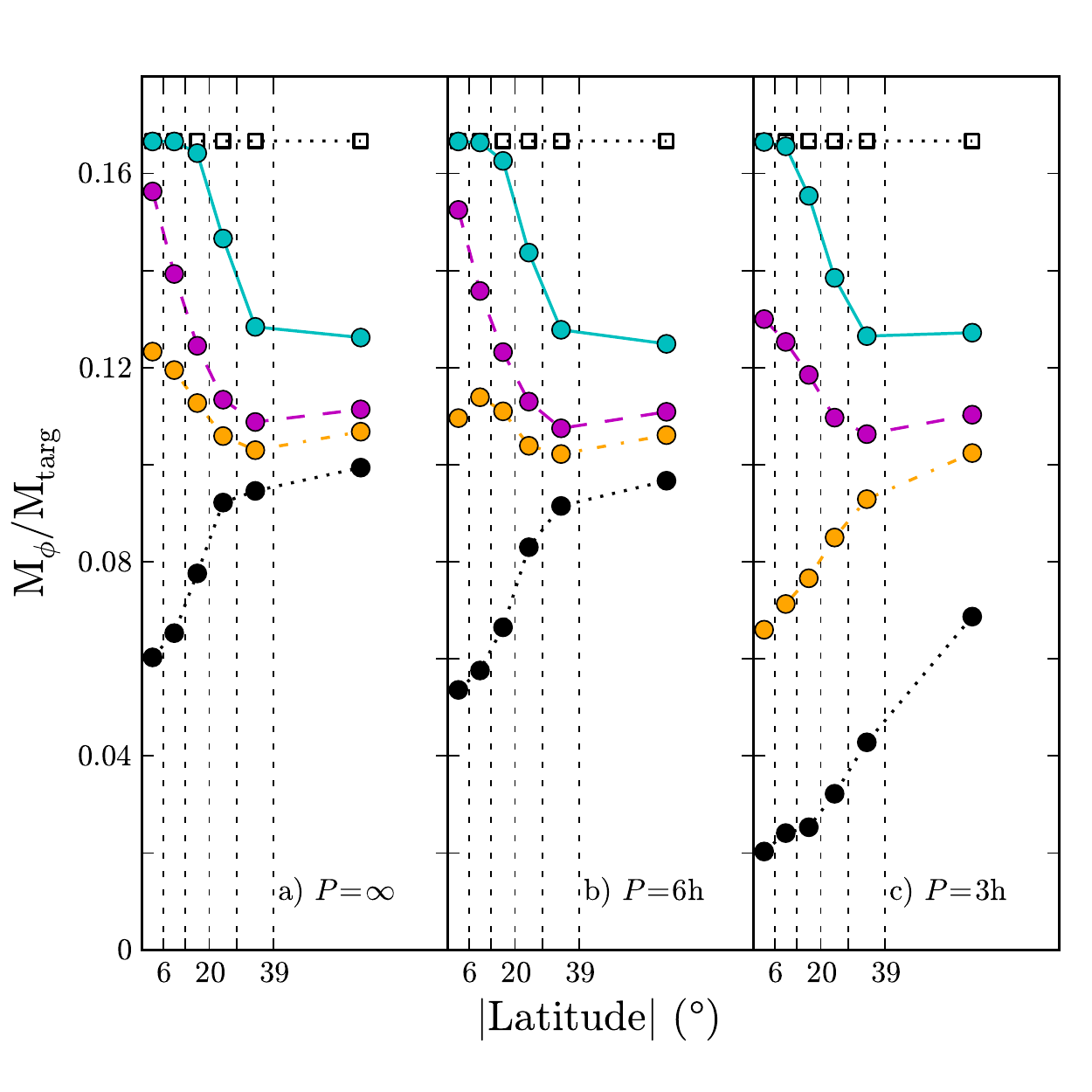} \
\caption{\footnotesize{Latitudinal origin of material from the target that constitutes the largest remnant for pole-on impacts ($\delta=0\degree$). $\mathrm{M_{\phi}}$ is the mass in the largest remnant originating from a given range in absolute latitude (six equal-mass bins) of the target. Solid cyan, dashed magenta, dot-dashed orange, and dotted black lines represent impact speeds of 5, 7, 8, and 10 m s$^{-1}$, respectively. The open black squares connected by the dotted lines represent the initial latitudinal mass distribution of the target. Panel (a) shows that, for a non-rotating target, higher impact speeds cause more mass to escape from the equator rather than the poles. As Panels (b) and (c) show, this effect is magnified for faster spin rates, leading to large enhancements in global mass loss, as shown in Figure 6.}}
\label{f:latitude_delta}
\end{figure}

\section{Discussion}

\subsection{The ``Universal'' Law for Catastrophic Disruption}

\indent In order to account for the dependence of mass ratio on catastrophic disruption criteria, Leinhardt \& Stewart (2009) introduced new variables into their formulation for predicting collision outcomes: the reduced mass $\mu \equiv M_{\mathrm{proj}}M_{\mathrm{targ}}/M_{\mathrm{tot}}$, the reduced-mass specific impact energy $Q_{R}\equiv0.5 \mu v^{2}_{\mathrm{imp}}/M_{\mathrm{tot}}$, and the corresponding reduced-mass catastrophic dispersal limit $Q^{\star}_{\mathrm{RD}}$. Through this new formulation, Leinhardt \& Stewart (2009) showed that the outcome of any head-on collision, regardless of projectile-to-target-mass ratio, can be described by a single equation that they call the ``universal'' law: 
\begin{equation}M_{\mathrm{LR}}/M_{\mathrm{tot}} = -0.5 (Q_{R}/Q'^{*}_{\mathrm{RD}})+0.5, \end{equation}
where the prime ($\prime$) notation in $Q^{\prime\star}_{\mathrm{RD}}$ was introduced by Leinhardt \& Stewart (2012) to denote a collision that could have a non-zero impact parameter. Leinhardt \& Stewart (2009) verified that, for head-on impacts ($b=0$), Equation (2) agrees well with results from both laboratory experiments and numerical simulations of binary collisions with a range of mass ratios and material properties. Leinhardt \& Stewart (2012) found that their numerical simulations showed deviations in $M_\mathrm{{LR}}/M_\mathrm{{tot}}$ of $\sim10\%$ for near-normal impacts ($b=0$ and $b=0.35$), and larger and more varied deviations for more oblique impacts.\\ 
\begin{figure*}[t!]
\centering
\includegraphics[height=2.5 in]{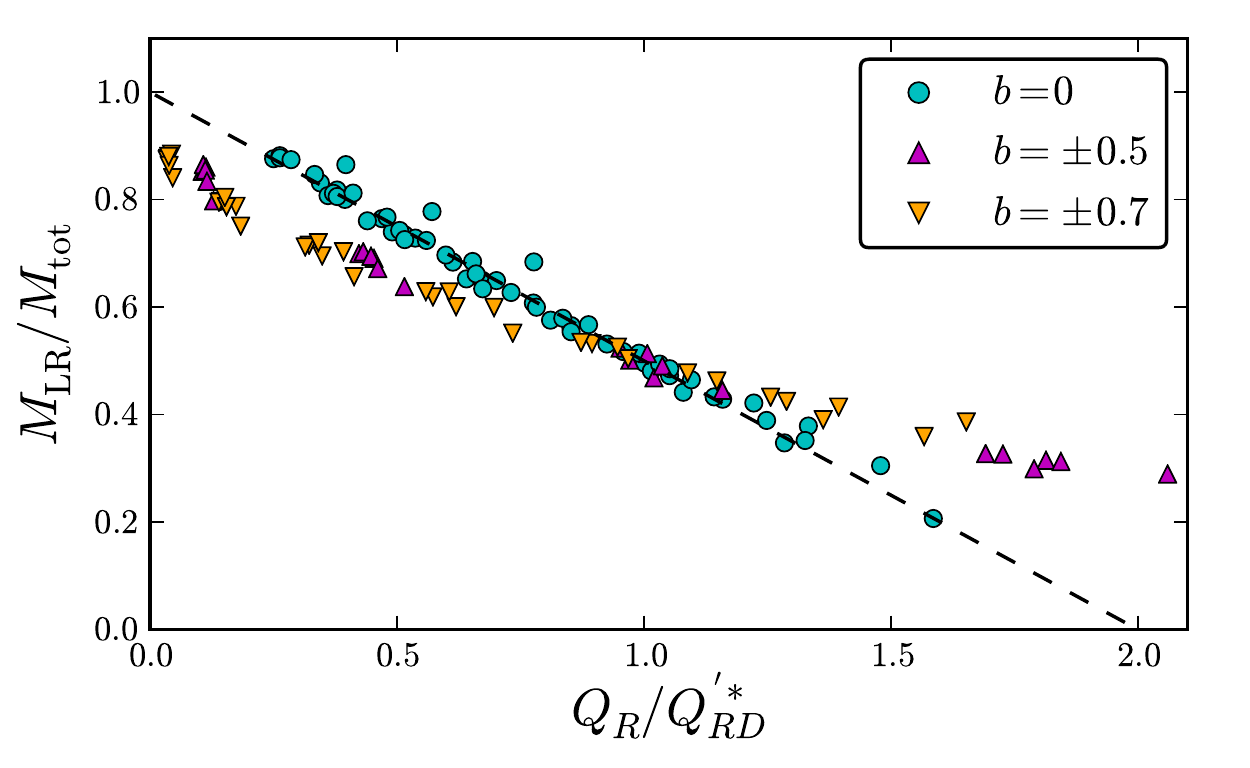} 
\caption{\footnotesize{Mass-loss outcomes for all simulations are plotted as a function of the specific impact energy in units of the catastrophic disruption criterion. Head-on collision outcomes are described well by the ``universal'' law for catastrophic disruption, Eq. (2) (dotted-line). Most head-on collisions (red circles) show $< 1 \%$ deviations from the universal law. Non-equatorial impacts ($\delta<90\degree$) onto rotating targets, show deviations of up to $\sim 10\%$ due to non-linear effects present when collisions are near- and super-catastrophic (see text for discussion). Oblique impacts (filled triangles) systematically deviate from the universal law.}}
\label{f:universal_law}
\end{figure*}
\indent In order to compare our results (summarized in Table 1) to those of Leinhardt \& Stewart (2012), the outcomes of each group of collisions ($M_\mathrm{{LR}}$ as a function of $Q_{R}$) were fit with a linear function to determine empirically the value of $Q^{\prime\star}_{\mathrm{RD}}$ (collision groups are uniquely identified by single values of $b$, $\delta$, and $P_\mathrm{{spin}}$.) For most collision scenarios the results are well described by the linear fit. In these cases, deviations from the model were less than $1\%$. For non-equatorial collisions ($\delta<90\degree$) onto a rotating target, we showed in Section $3.4$ the existence of two different mass loss outcome regimes. For near- and super-catastrophic collisions, enhanced mass dispersal from the equator causes a non-linear increase in mass loss. Hence, for these cases, a single linear fit to determine $Q^{\prime\star}_{\mathrm{RD}}$ leads to deviations of $\sim 10\%$.\\
\indent The results of our simulations are presented in Figure 8, superimposed on the universal law for catastrophic disruption. For collisions with $b=0$, we observe that the law predicts accurately the mass of the largest remnant (deviations $< 10\%$). Furthermore, varying the polar-impact angle, $\delta$, or the spin period, $P_\mathrm{{spin}}$, does not affect the mass of the largest remnant if the specific impact energy is normalized by $Q'^{*}_{\mathrm{RD}}$.\\
\indent Cases of $b=\pm 0.5, \pm 0.7$ seem to be better fit by a shallower slope. For oblique impacts, Leinhardt \& Stewart (2012) introduced a parameter $\alpha \equiv m_\mathrm{{int,proj}}/M_\mathrm{{proj}}$, which accounts for the mass of the projectile that interacts with the target, such that the appropriate reduced mass is
\begin{equation} \mu_{\alpha} \equiv \frac{\alpha M_{\mathrm{proj}}M_{\mathrm{targ}}}{\alpha M_{\mathrm{proj}}+M_{\mathrm{targ}} }. \end{equation}
\noindent For the case of a 1:10 mass ratio and $b=\pm 0.5$, we find $\alpha \sim 1.0$. Yet, we observe in our experiments that the specific impact energy required for catastrophic disruption is much greater than that for a head-on collision. Therefore, there must be some other mechanism that accounts for this discrepancy in required specific impact energy.\\
\indent During a collision, a compressive wave travels through the projectile, and upon encountering the projectile's edge, is reflected as a tensile wave that travels through the contact points and disrupts and disperses target (as well as projectile) material (Ryan 2000). The width of this wave depends on a number of physical parameters such as the strain loading rate, the type of material, and the size ratio of the projectile and target. We hypothesize that the size of this wave determines what fraction of the target interacts in the collision. For cases with non-zero $b$, we observe that the particles that are able to escape are mostly localized to the same hemisphere of the collision; hence, some of the material in the opposite hemisphere of the target shears off and is not involved in the collision. For head-on collisions, the dispersive wave would originate near the center and propagate symmetrically through the target, maximizing the amount of material affected by the collision. The exact wave mechanics involved in computing the fraction of the target that does interact during the collision is difficult to determine accurately analytically. For now, we attempt to find an empirical determination of this interacting target mass by adjusting the disruption criteria variables to include this effect. Hence, we introduce a further revised reduced mass:  
\begin{equation} \mu_{A} \equiv \frac{\alpha M_{\mathrm{proj}} A M_{\mathrm{targ}}}{\alpha M_{\mathrm{proj}}+ A M_{\mathrm{targ}} }, \end{equation}
\noindent where $A \equiv m_\mathrm{{int,targ}}/M_\mathrm{{targ}}$, the fraction of the mass of the target that interacts in the collision. Thus, the equivalent head-on specific impact energy is $Q^{A}_{R} = 0.5 \mu_{A} v^{2}_{\mathrm{imp}} / M_{\mathrm{tot}}$.  Furthermore, the mass of the largest remnant is now normalized by a total interacting mass, $M^{A}_{\mathrm{tot}}\equiv\alpha M_{\mathrm{proj}} + A M_{\mathrm{targ}}$. Using these new variables, we rescale the oblique impacts into equivalent head-on impacts. By using a $\chi^2$-minimization routine, we determined the best-fit values for $\alpha$ and $A$ (Figure 9) that rescale the oblique impact results such that they fit the universal law for catastrophic disruption. We consider two cases, $b=\pm0.5$ and $b=\pm0.7$. Since the effect is considered to be purely geometric, it is independent of the sign of $b$. For both cases of $b$, we find single $1/\chi^2$ peaked regions in the parameter space. Possible values of $A$ are well constrained between $0.8$ and $0.9$ for both cases of $b$. The best-fit values of $\alpha$ differs for the two $|b|$ cases, with $b =\pm0.5, \pm0.7$ peaked at $\alpha \sim 0.7, 0.75$, respectively.\\
\begin{figure}[h!]
\centering
\includegraphics[width=0.5\textwidth]{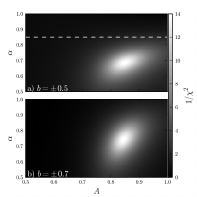} 
\caption{\footnotesize{Using a $\chi^2$ analysis, we determined the best values of $\alpha$ and $A$ that adjust the catastrophic disruption variables such that oblique impacts are well modeled by the universal law for catastrophic disruption. The white dashed line is the geometric constraint placed by $\alpha_{\mathrm{geom}}$. For $b = \pm 0.7$, $\alpha_{\mathrm{geom}}=0.44$. The $\chi^2$ analysis shows that the geometric model may be underestimating the fraction of interacting projectile material (see text for discussion). }}
\label{f:chi2}
\end{figure}
\indent In order to constrain the possible values of $\alpha$ and $A$, we also determined the interacting mass fraction of the projectile, which we will call $\alpha_{\mathrm{geom}}$, through a simple geometric model. Following Leinhardt \& Stewart (2012), we determine $\alpha_{\mathrm{geom}}$ by considering the volume of the projectile whose cross section intersects with the target. The volume of a spherical cap can be expressed as:
\begin{equation} V_\mathrm{{int,proj}} = \frac{\pi l^2}{3} (3r - l), \end{equation}
where $l$ is the height of the spherical cap, and $r$ is the radius of the sphere. For an oblique impact, $l$ is the projected length of the projectile overlapping the target, and can be expressed as
\begin{equation} l = (1-b)(R_{\mathrm{targ}}+R_{\mathrm{proj}}). \end{equation}
Therefore, 
\begin{equation} \alpha_{\mathrm{geom}} = \frac{m_\mathrm{{int,proj}}}{M_{\mathrm{proj}}} = \frac{V_{\mathrm{proj}}}{V_\mathrm{{int,proj}}} = \frac{3rl^2 - l^3}{4r^3}. \end{equation}
The value of $\alpha_{\mathrm{geom}}$ gives a lower boundary to the possible value of $\alpha$, since it is expected that the minimum amount of interacting material would be the mass that overlaps the target geometrically. We determined the best values of $\alpha$ and $A$ as constrained by the geometric model, and find that as $|b|$ increases, $\alpha$ and $A$ decrease (Table 2). This trend is expected, since more projectile and target material can shear off when the impact is close to grazing. For the cases with $b=\pm0.7$, we find that the best-fit value of $\alpha$ (0.745) is much greater than $\alpha_{\mathrm{geom}}$ (0.44). This suggests that the simple geometric model underestimates the value of $\alpha$. A fraction of the projectile whose cross section does not overlap with the target does not completely shear off; rather, it is involved in the collision process, contributing to the impact energy delivered to the target.\\ 
\indent For the cases with $b=\pm0.5$, we find that the best-fit values for $\alpha$ and $A$ are excluded when $\alpha$ is imposed to be greater than or equal to $\alpha_{\mathrm{geom}}$. If the geometric constraint is not considered, then a combinaton of $\alpha = 0.685$ and $A=0.845$ gives the minimum $\chi^2$. However, this would imply that the cases of $b=\pm0.5$ have a lower value for $\alpha$ than the cases of $b=\pm0.7$ ($\alpha=0.745$). This is unlikely, as a larger fraction of projectile material is expected to interact when a collision is closer to head-on.\\
\indent Hence, while our $\chi^2$ analysis does a good job of fitting the data to the universal law, the results are unphysical unless they are constrained by a geometric model. Therefore, more work must be done in order to verify whether the energetics of oblique impacts are affected by the interacting fractional mass of both the target and the projectile as we suggest here. In Figure 10, we show that oblique impacts follow the linear universal law if the axes are changed to our rescaled catastrophic disruption variables. However, we find that for super-catastrophic collisions, the data points seem to tail-off rather than follow a linear relationship. Laboratory experiments (Matsui et al.\ 1982, Kato et al.\ 1995) and disruption simulations (Korycansky \& Asphaug 2009, Leinhardt \& Stewart 2012), have shown that the mass of the largest remnant follows a power law with $Q_{R}$ for super-catastrophic collisions. At these high energies, mass dispersal results in the formation of a large number of fragments of roughly equal size, rather than a single large remnant, such that for incrementally higher impact energies, the largest remnant remains constant. This may explain the discrepancy that we see between the high-energy collision outcomes and the universal law. 
{\renewcommand{\arraystretch}{1.2}
\begin{table}[h]
  \centering
  \caption{\footnotesize{Summary of $\chi^2$ Analysis: Best-Fit Interacting Mass Fractions} }
  \label{t:alphaAchi2}
  \begin{tabular}{c|cccc}
    \hline
    \hline
    $b$ & $\alpha$ & $A$ & $\chi^2$ & $\alpha_{\mathrm{geom}}$\\
    \hline
    $\pm 0.5$ & 0.85 & 0.94 & 0.61 & 0.85\\
    $\pm 0.7$ & 0.745 & 0.845 & 0.68 & 0.44\\
    \hline
  \end{tabular}
\end{table}
}
\begin{figure}[h!]
\centering
\includegraphics[width=0.5\textwidth]{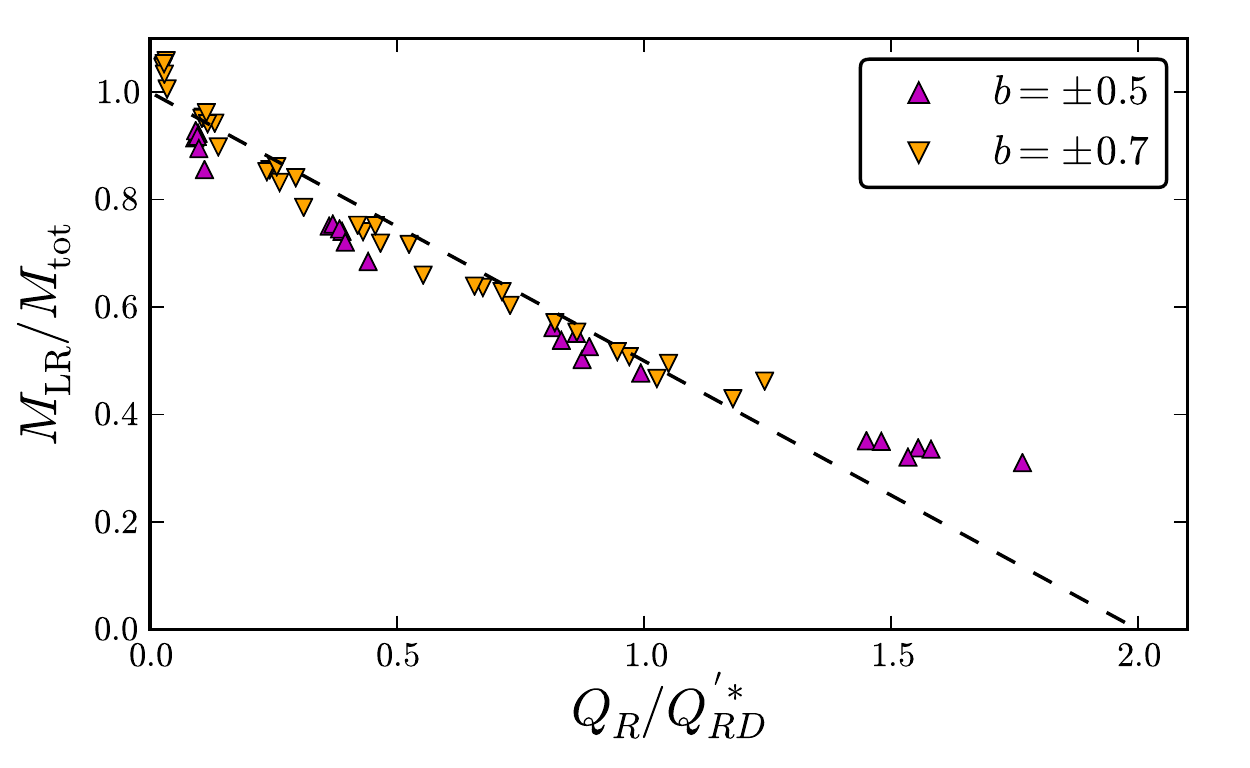} 
\caption{\footnotesize{Adjusting our results for the correct interacting projectile and target produces a better fit for oblique impacts, for low $Q^{A}_{R}/Q^{\star}_{\mathrm{RD}}$ (cf.\ Figure 8). The corrected interacting mass fractions are summarized in Table 2.}}
\label{f:universal_fix}
\end{figure}

\subsection{Rotation Dependence of Catastrophic Disruption for Head-on Equatorial Collisions}

Since rotation decreases the effective gravitational binding energy of a body, we first describe the size-dependence of catastrophic disruption so that we may be able to formulate an analytic description of the dependence on pre-impact rotation. The catastrophic disruption criterion is a function of radius, with two regimes: a strength-dominated regime and a gravity-dominated regime (Housen \& Holsapple 1990). For rocky bodies, the transition from strength to gravity occurs at a radius of $\sim 100$ m  (Leinhardt et al.\ 2008). In the strength regime, the catastrophic disruption criterion decreases with increasing radius; this is due to multiple factors, such as the increase in the size of the largest internal crack and the total number of flaws with target size. In the gravity regime, disruption increases as the radius increases since disruption requires shattering and gravitational dispersal, and the gravitational binding energy of a body, $U$, is proportional to the square of the body's radius. For a binary collision, the gravitational binding energy can be approximated as,\\ 
\begin{equation} U=\frac{3GM_{\mathrm{tot}}}{5R_{C1}}=\frac{4}{5}\pi\rho_{1}GR^{2}_{C1}, \end{equation}
where $G$ is the gravitational constant and $R_{C1}$ is the spherical radius of the combined projectile and target masses at a density of $\rho_{1}=1$ g cm$^{-3}$. Leinhardt \& Stewart (2009) introduced $R_{C1}$ in order to compare collisions of different projectile-to-target-mass ratios. \\
\indent By determining the dependence of mass ratio on the catastrophic disruption criterion, Leinhardt \& Stewart (2012) found that, in the gravity regime, the disruption criterion for equivalent equal-mass impacts, $Q^{\star}_{\mathrm{RD}}$, of different materials all fall along a single curve that scales as the radius squared. Since catastrophic disruption and the gravitational binding energy scale similarly with radius, the authors define a principal disruption curve, where $Q^{\star}_{\mathrm{RD}}$ is a scalar multiple of $U$. They defined a dimensionless material parameter, $c^{\star}$, that represents this offset, such that
\begin{equation}Q^{\star}_{\mathrm{RD}} = c^{\star} U .\end{equation}  
For bodies with a diverse range of material properties (strengthless hydrodynamic targets, rubble piles, ice, and strong rock targets) and with $0.5$ km $< R_{C1} < 1000$ km, Leinhardt \& Stewart (2012) found that the threshold for catastrophic disruption is defined by a single principal disruption curve with $c^{\star} = 5 \pm 2$. They find that a similar curve can describe the disruption of planet-size bodies ($R_{C1} > 3000$ km) with $c^{\star} = 1.9 \pm 0.3$.\\
\indent In order to formulate a description of the dependence of catastrophic disruption on rotation, we first consider a rotating fluid body that has an effective specific gravitational binding energy, $U_\mathrm{{eff}}$, given by
\begin{equation} U_\mathrm{{eff}} = U - \frac{|\bsym{\omega} \bsym{\cross} \bsym{r}|^{2}}{2}, \end{equation}
where $\bsym{\omega}$ is the constant angular velocity of the body, and $\bsym{r}$ is the position vector of a particle relative to the center of the target. Similarly, we propose that in a binary head-on equatorial collision with pre-impact rotation the catastrophic disruption criteria, $Q^{\star}_\mathrm{{RD,rot}}$, is a function of the angular speed of the target, $\omega_\mathrm{{targ}}$. For a head-on equatorial collision with pre-impact rotation (collision geometry with no dependence on $b$ and $\delta$), this is represented by a subtraction of a latitude-averaged centrifugal term as in Equation (10) such that
\begin{equation} Q^{\star}_\mathrm{{RD,rot}} = Q^{\star}_\mathrm{{RD,\norot}} - \omega^{2}_\mathrm{{targ}} R^{2}_\mathrm{{targ}}, \end{equation}
where $Q^{\star}_\mathrm{{RD,\norot}}$ is the catastrophic disruption criterion without pre-impact spin. For non-equatorial or oblique impacts, we expect that the change in the catastrophic disruption criterion is a more complicated function of collisional angular momentum and the efficiency of its transfer. For the purposes of this paper, we restrict our analysis to the simpler equatorial head-on collisions. Dividing Equation (11) by $Q^{\star}_\mathrm{{RD,\norot}}$, and substituting from Equation (9), we find 

\begin{eqnarray}\begin{split}
\frac{Q^{\star}_\mathrm{{RD,rot}}}{Q^{\star}_\mathrm{{RD,\norot}}} & = & 1 - K \left(\frac{\omega_\mathrm{targ}}{\omega_\mathrm{crit}}\right)^{2};\\
K & \equiv & \frac{5}{3}\frac{R_{C1}}{R_\mathrm{{targ}}}\frac{M_\mathrm{{targ}}}{M_\mathrm{{tot}}}\frac{1}{c^{\star}},
\end{split}\end{eqnarray}

\noindent where $\omega_\mathrm{{crit}}\equiv(GM_\mathrm{{targ}}/R^{3}_\mathrm{{targ}})^{1/2}$ is the spin break-up limit of the target. We fit Equation (12) with the empirically derived catastrophic disruption values (described in Section 4.1) for the three different pre-impact spin cases normalized by the spinless case (Figure 11). We find a best-fit value of $K = 0.3814$ (maximum deviations were $\sim 3\%$), corresponding to a value of $c^{\star} = 4.83$. Our value of $c^{\star}$ falls within the range for small bodies ($R<1000$ km in size) that Leinhardt \& Stewart (2012) find.\\ 
\indent Equation (12) predicts that for pre-impact rotations close to break-up, the catastrophic disruption criterion can decrease by a factor of $\sim 40\%$. In reality, a spherical target spinning close to break-up would reach a new fluid-equilibrium shape, and the nature of its ellipsoidal shape will likely affect the catastrophic disruption criterion. In our simulations, the impacts occur quickly enough that the target does not reach a fluid equilibrium before disrupting. Future work will study the dependence of catastrophic disruption on the pre-impact shape of the target body. 

\begin{figure}[h!]
\centering
\includegraphics[width=0.5\textwidth]{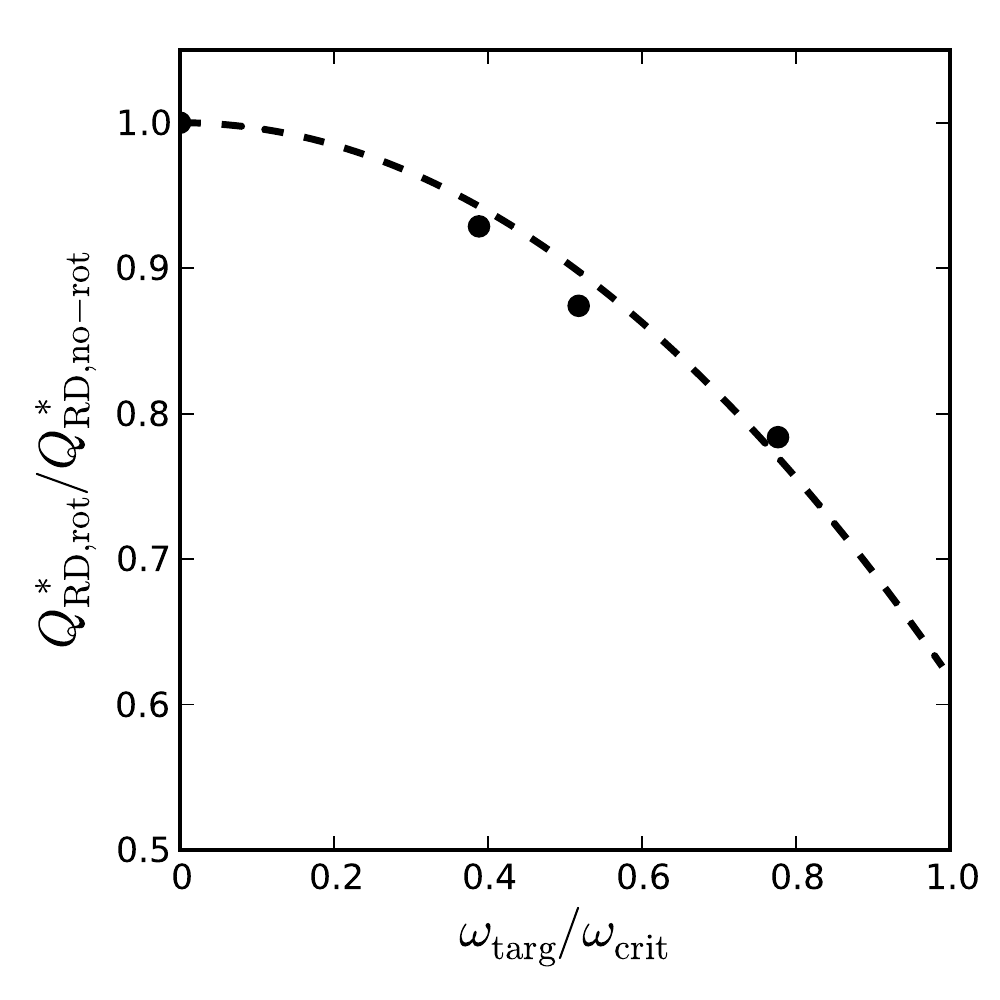} 
\caption{\footnotesize{Head-on equatorial impacts have catastrophic disruption thresholds that are sensitive to pre-impact spin. We develop a semi-analytic description of $Q^{\star}_{\mathrm{RD}}$ as a function of the pre-impact spin-rate of the target, $\omega_\mathrm{{targ}}$ (see text), and find that our data imply a value of $c^{\star}=4.83$ for the dimensionless material parameter of the principal disruption curve. The dashed curve is the best-fit function (of the form described by Equation (12)) to the catastrophic disruption thresholds with pre-impact rotation normalized by the case with no rotation (circles). }}
\label{f:omega}
\end{figure}

\section{Conclusions and Future Work}
We have studied the effect of initial rotation on the outcome of rubble-pile collisions by analyzing the properties of the largest remnant and material that is gravitationally bound to it. By simulating different collision geometries and speeds, we have begun to explore a parameter space that is wide enough that we can formulate a phenomenological description of collision outcomes. Our main conclusion is that mass dispersal is a function of initial rotation period, with faster-rotating rubble-pile targets dispersing more mass. By analyzing the initial spatial distribution of the gravitationally bound masses, we have shown that there is an enhancement of mass loss when the impact energy is efficiently transferred to the prograde hemisphere of a rotating rubble pile. For head-on impacts onto regions near the pole, the collision efficiently disperses equatorial material for near- and super-catastrophic collisions, and fast rotation increases mass loss by factors of up to $50\%$. The mass of the largest remnant of head-on impacts is well described by the ``universal law'' for catastrophic disruption first put forward by Leinhardt \& Stewart (2009), independent of initial pre-impact rotation. Hence, for a given impact speed, pre-impact rotation decreases $M_\mathrm{LR}/M_\mathrm{tot}$, and the corresponding decrease in $Q^{\prime\star}_{\mathrm{RD}}$ can be described by the universal law. However, oblique impacts follow a linear relationship with a shallower slope than the universal law. When the interacting mass fraction of the projectile and target are factored into the catastrophic disruption variables, the outcomes for oblique impacts can be rescaled for a better match to the universal law.\\
\indent By subtracing a centrifugal term from the catastrophic disruption criterion of the case with no pre-impact rotation, we developed a prescription that describes the change in the catastrophic disruption criteria of head-on equatorial impacts onto a rotating target. We independently find a dimensionless material parameter, $c^{\star}=4.83$, that agrees with the principal disruption curves of Leinhardt \& Stewart (2012) for small bodies. Our simplified description does not take into account the effects of angular momentum transfer in oblique impacts or the mechanism for enhanced mass loss from near- and super-catastrophic polar impacts described in Section 3.3. These effects will have to be further studied so that the change in the catastrophic disruption criterion can be determined for any impact trajectory.\\ 
\indent In the future, we will explore a wider parameter space in order to strengthen the conclusions drawn here. In particular, the effectiveness of rotation in enhancing mass loss must be studied for different projectile-to-target-mass ratios. Since mass loss is sensitive to rotation, the spin-up or spin-down of the post-impact largest remnant plays an important role in the size evolution of a population of km-size bodies. Furthermore, the change in spin likely has an effect on the reaccumulation process, changing the final mass. In some cases, spin-up may lead to a remnant crossing the rotational disruption threshold ($P_\mathrm{{spin}} \sim 2.3$ hr, for $\rho \sim 2$ g cm$^{-3}$). In order to understand how rotation affects the long-term size, shape, and spin evolution of a population of SSSBs, a semi-analytic description of the dependence of $Q^{\prime\star}_{\mathrm{RD}}$ on the parameters explored here must be formulated, such that $M_{\mathrm{LR}}$ can be determined for any given collision. Therefore, future work will also need to study the sensitivity of mass loss on rotation for larger ($> 1$ km) bodies. It is unclear whether the collisional dynamics explored here scales to larger bodies. This will help inform future planet-formation studies by giving a more accurate prescription for collision outcomes, a necessary component for models that study the collisional growth of planetesimals.\\
\indent Lastly, our study focused on a single set of SSDEM parameters, corresponding to a single type of material. However, there is a diversity of asteroid types, and their exact material properties are uncertain. Future space missions, in particular sample-return ones such as Hayabusa 2 (JAXA) to be launched in 2014--2015, and OSIRIS-REx (NASA) to be launched in 2016, will shed some light on the physical and dynamical properties of asteroids that will help constrain the plausible SSDEM values.  Until then, we are able to perform simulations that vary SSDEM values so that we can explore the range of possible outcomes. Furthermore, the SSDEM code is capable of simulating gravitational aggregates made up of a size distribution of particles. Walsh \& Richardson (2008) and Walsh et al.\ (2012) found that the Mohr--Coulomb internal angle of friction ($\phi$), effectively a rubble pile's shear strength, depends on the particle size distribution. Therefore, in order to accurately describe collisional processes in the solar system, future work will explore the dependence of impact outcomes on the different possible combinations of material properties of asteroids.
\section{Acknowledgements}
The authors would like to thank the referee, Dr. Keiji Ohtsuki, for his insightful and practical comments during the revision process. The authors would also like to acknowledge the initial work of student Sarah Munyan, and acknowledge us of which motivated this study. Simulations were performed on the YORP cluster administered by the Center for Theory and Computation, part of the Department of Astronomy at the University of Maryland. Part of the research leading to these results has received funding from the European Union Seventh Framework Programme (FP7/2007-2013) under grant agreement No 282703--NEOShield. This material is based on work supported by the U.S.\ National Aeronautics and Space Administration under grant Nos. NNX08AM39G, NNX10AQ01G, and NNX12AG29G issued through the Office of Space Science and by the U.S.\ National Science Foundation under grant No. AST1009579.

\end{document}